\documentclass{article}
\usepackage{euscript,amsmath,amssymb,amsfonts,graphicx,bm,amsthm,mathtools}
\usepackage{array}
\usepackage{cite}

  \usepackage{paralist}
  \usepackage{graphics} 
\usepackage[caption=false]{subfig}
\usepackage{soul}

\usepackage{latexsym,amssymb,enumerate,amsmath,amsthm} 
  \usepackage{graphicx} 
  \usepackage{tikz}
     \usepackage[colorlinks=false]{hyperref}
 \hypersetup{urlcolor=blue, citecolor=red}
\usepackage{latexsym,amssymb,enumerate,amsmath} 
\usepackage{pgfplots}
\usepackage{soul,xcolor}

\newcommand{\dd}{\text{d}}
\def\eps{\varepsilon}

\def\R{\mathbb{R}}

\def\c{\overline{c}}
\def\u{\overline{u}}
\def\r{\overline{r}}
\def\t{\overline{t}}

\newcommand{\Markov}[2]{\underset{#1}{\overset{#2}{\rightleftharpoons}}}
\def\ka{k_{\textup{a}}}
\def\kb{k_{\textup{b}}}
\def\kc{k_{\textup{c}}}
\def\kr{k_{\textup{r}}}
\def\ss{\textup{ss}}

\def\chib{\chi_{\textup{b}}}
\def\chic{\chi_{\textup{c}}}
\def\Jmax{J_{\textup{max}}}
\def\Jbp{J_{\textup{bp}}}
\def\Jmm{J_{\textup{mm}}}
\def\J{J_{*}}
\def\Vmax{V_{\textup{max}}}
\def\KM{K_{\textup{M}}}
\def\N{N}
\def\pt{\frac{\partial}{\partial t}}
\def\pr{\frac{\partial}{\partial r}}
\def\pz{\frac{\partial}{\partial z}}
\def\dt{\frac{\dd}{\dd t}}
\def\dr{\frac{\dd}{\dd r}}
\def\Nhalf{N_{*}}

\title{Revising Berg-Purcell for finite receptor kinetics}

\author{Gregory Handy\thanks{Departments of Neurobiology and Statistics, University of Chicago, Chicago, IL 60637 USA (\texttt{ghandy@uchicago.edu}).} \thanks{Grossman Center for Quantitative Biology and Human Behavior, University of Chicago, Chicago, IL, USA.} \and Sean D. Lawley\thanks{Department of Mathematics, University of Utah, Salt Lake City, UT 84112 USA (\texttt{lawley@math.utah.edu}). The second author was supported by the National Science Foundation (Grant Nos.\ 1944574, DMS-1814832, and DMS-1148230).}
}

\begin{document}
\maketitle

\begin{abstract}
From nutrient uptake, to chemoreception, to synaptic transmission, many systems in cell biology depend on molecules diffusing and binding to membrane receptors. Mathematical analysis of such systems often neglects the fact that receptors process molecules at finite kinetic rates. A key example is the celebrated formula of Berg and Purcell for the rate that cell surface receptors capture extracellular molecules. Indeed, this influential result is only valid if receptors transport molecules through the cell wall at a rate much faster than molecules arrive at receptors. From a mathematical perspective, ignoring receptor kinetics is convenient because it makes the diffusing molecules independent. In contrast, including receptor kinetics introduces correlations between the diffusing molecules since, for example, bound receptors may be temporarily blocked from binding additional molecules. In this work, we present a modeling framework for coupling bulk diffusion to surface receptors with finite kinetic rates. The framework uses boundary homogenization to couple the diffusion equation to nonlinear ordinary differential equations on the boundary. We use this framework to derive an explicit formula for the cellular uptake rate and show that the analysis of Berg and Purcell significantly overestimates uptake in some typical biophysical scenarios. We confirm our analysis by numerical simulations of a many particle stochastic system.
\end{abstract}

\section*{Introduction}

Many biological systems depend on molecules diffusing and interacting with membrane receptors. For example, cellular nutrient uptake relies on cell surface receptors binding and transporting diffusing molecules into the cell \cite{johnston1999}. Chemoreception and chemotaxis similarly depend on cell surface receptors binding extracellular diffusing molecules \cite{berg1977}. An important part of the immune response involves antibodies binding to epitopes on the surface of a virion \cite{sela2013, perelson1997}. In addition, synaptic transmission requires neurotransmitter molecules released from one neuron to diffuse across the synaptic cleft and bind to receptors on the adjacent neuron \cite{deutch2013}. 

\begin{figure}[t]%[t]%[htp]
\centering
\includegraphics[width=.99\linewidth]{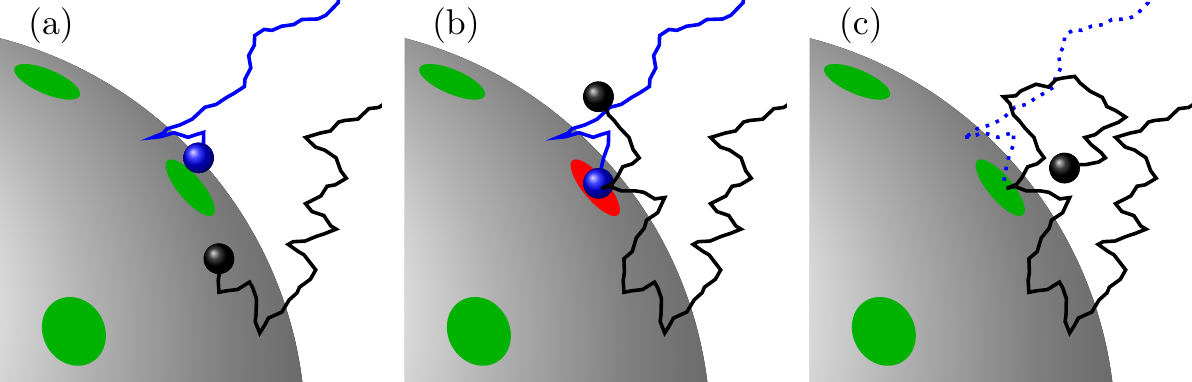}
\caption{(a) The blue and black molecules compete to bind to a surface receptor. (b) The blue molecule binds to a receptor. The black molecule is then temporarily blocked from binding to that receptor. (c) After some time, the blue molecule is absorbed by the receptor and that receptor is free to bind the black molecule.}
\label{figschem}
\end{figure}

In most instances, the receptors cannot bind molecules continuously, but rather binding one or more molecules temporarily hinders binding additional molecules (see Fig.~\ref{figschem} for an illustration). This could be due simply to mutual exclusion at the receptor (i.e.\ a receptor can only bind one molecule at a time). Alternatively, this effect could be due to the finite rate that a receptor can transport bound molecules into the cell, as in the case of nutrient transport \cite{fiksen2013, aksnes2011}. Similarly, the effect could stem from a finite receptor internalization rate \cite{aquino2010, ferguson2001, mukherjee1997}. In the case of synaptic transmission, a receptor that captures a molecule changes conformation, and during this time it cannot capture additional molecules \cite{deutch2013, handy2018, handy2019}. That is, the receptor must wait a transitory ``recharge'' time following the capture of any molecule before additional captures. A similar recharge time affects experiments where molecules are released into extracellular space in the brain and bind to receptors on astrocytes \cite{handy2017}. In ecology, this notion of a recharge time is called the handling time, and it is the time a predator (the ``receptor'') must wait after capturing a prey (the ``molecule'') before it can hunt again.

%%%%%%%%%%%%%%%%%%%%%
%%%%%%%%%%%%%%%%%%%%%

The common feature of these examples is that the receptors process molecules at finite kinetic rates. Mathematical models often neglect receptor kinetics, which greatly simplifies the analysis since it allows the diffusing molecules to be independent. Including receptor kinetics makes the diffusing molecules dependent, since the molecules may affect each other through their interactions with the receptors. For example, if one molecule binds to a receptor, then additional molecules may be temporarily blocked from that receptor. 

An important example of mathematical analysis that neglects receptor kinetics is the formula of Berg and Purcell for the rate that cell surface receptors capture extracellular molecules \cite{berg1977}. Assuming that the receptors occupy a small fraction of the cell surface, they found that the uptake rate is 
\begin{align}\label{jbp0}
\Jbp
:=\frac{\eps {\N}}{\eps {\N}+\pi}\Jmax,
\end{align}
where $\eps$ is the ratio of the receptor radius to the cell radius, ${\N}\ge1$ is the number of receptors, and $\Jmax$ is the uptake rate if the entire surface is covered by perfectly absorbing receptors \cite{berg1977}. In particular, $\Jmax$ is \cite{smoluchowski1917}
\begin{align}\label{jmax0}
\Jmax
:=4\pi DRu_{0},
\end{align}
where $R$ is the cell radius, $D$ is the diffusivity of extracellular molecules, and $u_{0}$ is the concentration of extracellular molecules.

Berg and Purcell's formula in Eq.~\ref{jbp0}  has been very influential. Indeed, many works have sought to refine equation~Eq.~\ref{jbp0} to incorporate the effects of other details in the problem. For example, Zwanzig used an effective medium formalism to account for the effects of interference between receptors \cite{zwanzig1990}. Other works have modified Eq.~\ref{jbp0} to include other effects, including receptor arrangement, cell membrane curvature, and receptor motion \cite{zwanzig1991, bernoff2018b, lindsay2017,Berezhkovskii2004,Berezhkovskii2006,Dagdug2016,Eun2017,Muratov2008,eun2020,lawley2019bp,kaye2019}. In one particularly important study, Wagner et al.\ \cite{wagner2006} extended Eq.~\ref{jbp0} to non-spherical geometries and used this analysis to argue that the cylindrical morphology of cell envelope extensions serves to increase nutrient uptake. 

To derive Eq.~\ref{jbp0}, Berg and Purcell \cite{berg1977} assumed that any molecule that touches a receptor 
\begin{quote}
``is immediately (or within a time short compared to the interval between arrivals) captured and transported through the cell wall, clearing the site for its next catch.''
\end{quote}
This assumption makes the diffusing molecules independent. However, it is clear that this assumption is violated at sufficiently high extracellular concentrations.

In this paper, we present a modeling framework for coupling bulk diffusion of molecular species to surface receptors with finite kinetics. Mathematically, this framework uses boundary homogenization to link the diffusion equation (a partial differential equation (PDE)) to boundary conditions described by nonlinear ordinary differential equations (ODEs). We then reduce this PDE-ODE system to a PDE with a nonlinear boundary condition of Michaelis-Menten type. We confirm the predictions of this framework and analysis with detailed numerical simulations of a full, many particle stochastic system. While the general framework can be applied in a variety of problems and geometries, we develop the theory primarily in the context of the Berg-Purcell problem described above. We derive an explicit formula for the cellular uptake rate as a function of the various parameters in the problem, including the kinetic rates of receptors. We show that the classical result in Eq.~\ref{jbp0} significantly overestimates uptake in some typical biophysical scenarios. 

The rest of the paper is organized as follows. We first rederive the classical result in Eq.~\ref{jbp0}, and then we present the PDE-ODE framework and derive a reduced Michaelis-Menten boundary condition. Next, we use the modeling framework to find explicit formulas for various steady-state quantities, including cellular uptake and the fraction of bound receptors. We then describe our numerical methods and verify the predictions of the modeling framework by stochastic simulations. Finally, we explore some biophysical implications of our results in typical parameter regimes of interest. We conclude by discussing related work and highlighting future directions. 

\section*{Methods}

\subsection*{Berg-Purcell and boundary homogenization}\label{bpreview}

We begin by reviewing the model of Berg and Purcell \cite{berg1977} and boundary homogenization \cite{shoup1982, bernoff2018b, Berezhkovskii2004, Muratov2008, lawley2019fpk, plunkett2020}. Consider a spherical cell in a large medium containing spherical molecules of some substrate. Fixing our reference frame on the cell, the substrate concentration, $u=u(r,\theta,\varphi,t)$, satisfies the diffusion equation in spherical coordinates $(r,\theta,\varphi)$,
\begin{align}\label{pde11}
{\pt}u=D\Delta u,\quad r>R,
\end{align}
where $D>0$ is the substrate diffusivity and 
$R>0$ is the cell radius. The concentration is held constant far from the cell, 
\begin{align}\label{ffc}
\lim_{r\to\infty}u=u_{0}>0.
\end{align}

The cell has ${\N}\gg1$ surface receptors for the substrate. The receptors are roughly evenly distributed on the cell surface, and each receptor is a small circular patch of radius $\eps R$ with $\eps\ll1$. Substrate molecules can be absorbed by receptors and otherwise reflect from the cell surface. We thus have mixed boundary conditions at $r=R$,
\begin{align}\label{bc}
\begin{split}
D{\pr}u&=\frac{D}{R}\kappa_{\text{rec}}u,\, \,r=R,\text{ and }(\theta,\varphi)\text{ in a receptor},\\
{\pr}u&=0,\quad\,\, r=R,\text{ and }(\theta,\varphi)\text{ not in a receptor},
\end{split}
\end{align}
where 
\begin{align}\label{kapparec}
\kappa_{\text{rec}}
\in(0,\infty)\cup\{\infty\}
\end{align}
is a dimensionless parameter describing the rate that a receptor binds a substrate molecule. Berg and Purcell took $\kappa_{\text{rec}}=\infty$, which means that a substrate is immediately absorbed upon contact with a receptor \cite{berg1977}. If $\kappa_{\text{rec}}=\infty$, then the first boundary condition in Eq.~\ref{bc} means $u=0$.

The method of boundary homogenization replaces the heterogeneous boundary conditions in Eq.~\ref{bc} by a homogeneous boundary condition of the form,
\begin{align}\label{replace}
D{\pr}u&=\frac{D}{R}\kappa u,\quad r=R,
\end{align}
for some dimensionless trapping rate $\kappa>0$. The trapping rate $\kappa$ is chosen to encapsulate the effective binding properties of the heterogeneous surface in Eq.~\ref{bc}. Notice that $\kappa\ll1$ corresponds to a surface which is mostly reflecting, whereas $\kappa\gg1$ corresponds to a surface which is mostly absorbing. The benefit of boundary homogenization is that it makes the substrate concentration $u=u(r,t)$ depend only on the radius $r\ge R$ and time $t\ge0$, and not on the angular variables $(\theta,\varphi)$ (assuming the initial condition is independent of $(\theta,\varphi)$).

The idea behind boundary homogenization is that, due to diffusion in the angular variables, the surface heterogeneity only affects the substrate concentration near the surface. In particular, the concentration is constant in the angular variables outside a boundary layer, where the width of this layer depends on the length scale of the surface heterogeneity. Many sophisticated methods have been developed to choose the trapping rate $\kappa$ in Eq.~\ref{replace} in order to incorporate the number, size, and arrangement of receptors \cite{bernoff2018b, lindsay2017,Berezhkovskii2004,Berezhkovskii2006,Dagdug2016,Eun2017,Muratov2008,eun2020,lawley2019bp,kaye2019}. If the receptors occupy a small fraction of the cell surface, then the trapping rate is linear in the number of receptors ${\N}$ and is given by \cite{shoup1982, zwanzig1991, Berezhkovskii2004} 
\begin{align}\label{trapping}
\kappa
=\frac{\eps {\N}}{\pi }\Big(1+\frac{4}{\eps \pi \kappa_{\text{rec}}}\Big)^{-1},
\end{align}
where $1/\kappa_{\text{rec}}=0$ if $\kappa_{\text{rec}}=\infty$.

It is straightforward to solve Eqs.~\ref{pde11}-\ref{ffc} and Eq.~\ref{replace} at steady-state to obtain the large-time substrate flux into the cell by integrating over the surface of the sphere of radius $R>0$ \cite{berg1977}, 
\begin{align}\label{bpg}
\lim_{t\to\infty}D\int_{r=R}{\pr}u\,\dd S
=\frac{\kappa }{\kappa +1}\Jmax,
\end{align}
where $\Jmax$ is in Eq.~\ref{jmax0} and is the flux in the case that the entire cell surface is absorbing \cite{smoluchowski1917}. If $\kappa_{\text{rec}}=\infty$, then Eq.~\ref{trapping} and Eq.~\ref{bpg} yield the Berg-Purcell \cite{berg1977} flux formula in Eq.~\ref{jbp0}.

%%%%%%%%%%%%%%%%%%%%%%%%%%%%%%%%%%%%%%%%%%%%%%%%%%%%%%%%%%%%%%%%%%%%%%%%%%%%%%%%%%%%%%%%%%%%%%%%%%%%%%%%%%%%
\subsection*{Including finite receptor kinetics}\label{framework}

The model above assumes that receptors can continuously absorb substrates. That is, it assumes that there is no limit to the rate that a receptor can process substrate molecules. This modeling assumption is valid if receptors process molecules much faster than molecules tend to hit receptors. 

When is a system in this parameter regime? If we use the following characteristic values (used in, for example, \cite{berg1977}),
\begin{align*}
D
=10^{3}\,\mu\text{m}^{2}\text{s}^{-1},\quad
R
=1\,\mu\text{m},\quad
u_{0}
=1\,\mu\text{M},
%=6\times10^{2}\,\#\mu\text{m}^{-3},
\end{align*}
then molecules arrive to the cell surface at rate (see Eq.~\ref{jmax0})
\begin{align*}
\Jmax
%=4\pi DRu_{0}
%=4\pi(10^{3}\,\mu\text{m}^{2}\text{s}^{-1})(1\,\mu\text{m})(6\times10^{2}\mu\text{m}^{-3})
=7.5\times10^{6}\,\text{s}^{-1}.
\end{align*}
If we use the following characteristic values (again, see \cite{berg1977}) for the dimensionless receptor radius $\eps$ and the number of receptors ${\N}$,
\begin{align*}
\eps
=10^{-3},\quad {\N}=10^{3},
\end{align*}
then Eq.~\ref{jbp0} implies that the arrival rate to a single receptor is
\begin{align}\label{est}
\frac{\Jbp}{{\N}}
=\frac{\eps}{\eps {\N}+\pi}\Jmax
=1.8\times10^{3}\,\text{s}^{-1}.
\end{align}

Hence, in this parameter regime, the Berg-Purcell formula in Eq.~\ref{jbp0} gives a valid estimate of cellular uptake if a single receptor can transport molecules through the cell wall at a rate much faster than Eq.~\ref{est}. However, the so-called turnover rates of membrane transporters are usually in the range \cite{bionumbers}
\begin{align*}
\kc\in[3\times10^{1},3\times10^{2}]\,\text{s}^{-1},
\end{align*}
which is much slower than Eq.~\ref{est}.

%%%%%%%%%%%%%%%%%%%%%%%%%%%%%%%%%%%%%%%%%%%%%%%%%%%%%%%%%%%%%%%%%%%%%%%%%%%%%%%%%
\subsubsection*{Receptors modeled by ODEs}

To model membrane receptors with finite kinetics, we suppose that the substrate molecules interact with the receptors via a classical substrate and enzyme reaction scheme, 
\begin{align}\label{reaction}
S+E\Markov{{\kb}}{{\ka}}C\xrightarrow{{\kc}} P+E.
\end{align}
Here, $\ka>0$, $\kb\ge0$, and $\kc\ge0$ denote respectively the rates of ``association'', ``breakup'', and ``catalysis'' \cite{gunawardena2007}. In this formalism, freely diffusing extracellular substrates are represented by $S$, and the receptors play the role of the enzyme $E$. In particular, a substrate $S$ diffuses and binds to a receptor $E$ when it forms the complex $C$. During this time, the receptor is unavailable to bind with another substrate molecule, until it ``recharges'' by either producing $P$ or releasing the substrate $S$. The product $P$ could represent a substrate molecule that was transported into the cell, or a modified substrate molecule that was released back into the extracellular bulk but can no longer interact with receptors. Note that we allow for the possibility that $\kb=0$ or $\kc=0$. 

Mathematically, we replace the homogenized boundary condition in Eq.~\ref{replace} at the cell surface by
\begin{align}\label{bc11}
D{\pr}u(R,t)
&={\ka}e(t) u(R,t)-{\kb}c( t),
\end{align}
where the free receptor and bound receptor concentrations, $e(t)$ and $c(t)$, satisfy the ODEs,
\begin{align}
\begin{split}\label{ode}
	{\dt}e(t) &=-{\ka}e(t)u(R, t)+({\kb}+{\kc})c(t), \\
	{\dt}c(t) &={\ka}e(t)u(R, t)-({\kb}+{\kc})c(t).
	\end{split}
\end{align}
Adding the equations in Eq.~\ref{ode}, we see that the total receptor concentration, given by the number of receptors per the surface area of the cell, 
\begin{align}\label{e0}
e_{0}
:=\frac{{\N}}{4\pi R^{2}},
\end{align}
is conserved. That is, $e(t)=e_{0}-c(t)$. Note that $\ka$ has dimension $[\ka]=(\text{length})^{3}(\text{time})^{-1}$, whereas $\kb$ and $\kc$ are pure rates with dimension $[\kb]=[\kc]=(\text{time})^{-1}$. Further, if all of the receptors are free (i.e.\ $c(t)=0$), we choose $\ka$ so that Eq.~\ref{bc11} reduces to Eq.~\ref{replace}. In particular, we take
\begin{align}\label{ka}
\ka
=\frac{\kappa D}{e_{0}R}
%=\frac{\kappa 4\pi DR}{m}
%=\frac{Dm\eps 4\pi R^{2}}{m\pi R}\Big(1+\frac{4D}{\pi\eps R}\frac{1}{\kappa_{\text{rec}}}\Big)^{-1}
=4 D\eps R\Big(1+\frac{4}{\eps \pi \kappa_{\text{rec}}}\Big)^{-1}>0.
\end{align}
Summarizing, the model consists of the following PDE with nonlinear coupling to an ODE through a boundary condition,
\begin{align}\label{summary}
\begin{split}
{\pt}u
&=D\Delta%\Big(\frac{2}{r}{\pr}+\frac{\partial^{2}}{\partial r^{2}}\Big)
u,\quad r>R,\,t>0,\\
\lim_{r\to\infty}u(r,t)
&=u_{0}>0,\\
D{\pr}u(R,t)
&={\ka}\big(e_{0}-c(t)\big) u(R,t)-{\kb}c( t),\\
{\dt}c(t)
&={\ka}\big(e_{0}-c(t)\big)u(R, t)-({\kb}+{\kc})c(t).
\end{split}
\end{align}

%%%%%%%%%%%%%%%%%%%%%%%%%%%%%%%%%%%%%%%%%%%%%%%%%%%%%%%%%%%%%%%%%%%%%%%%%%%%%%%%%%%%%%%%%%%%%%%%%%%%%%%%%%%%
\subsubsection*{Michaelis-Menten boundary condition}

Defining the dimensionless variables,
\begin{align*}
\r
=\frac{r}{R},\quad
\t
=\frac{t}{R^{2}/D},\quad
\u
=\frac{u}{u_{0}},\quad
\c
=\frac{c}{e_{0}},
\end{align*}
Eq.~\ref{summary} becomes
\begin{align}
\frac{\partial}{\partial \t}\u
&=\Delta%\Big(\frac{2}{\r}\frac{\partial}{\partial \r}+\frac{\partial^{2}}{\partial \r^{2}}\Big)
\u,\quad \r>1,\,\t>0,\nonumber\\
\lim_{\r\to\infty}\u(\r,\t)
&=1,\nonumber\\
\frac{\partial}{\partial \r}\u(1,\t)
&={{{\kappa}}}\big(1-\c(\t)\big)\u(1,\t)-\chi_{\text{b}}\c(\t),\nonumber\\
\frac{\delta}{{{{\kappa}}}}\frac{\dd}{\dd \t}c(\t)
&=\u(1, \t)-\Big(\frac{\chi_{\text{b}}+\chi_{\text{c}}}{{{{\kappa}}}}+\u(1, \t)\Big)\c(\t),\label{mm}
\end{align}
where $\kappa$ is in Eq.~\ref{trapping} and 
\begin{align}\label{dp}
%{{{\kappa}}}
%:=\frac{\eps {\N}}{\pi}\Big(1+\frac{4}{\eps \pi \kappa_{\text{rec}}}\Big)^{-1},\quad
\delta
:=\frac{{\N}}{4\pi R^{3}}\frac{1}{u_{0}},\quad
\chi_{\text{b}}
:=\frac{{\N}\kb}{\Jmax},\quad
\chi_{\text{c}}
:=\frac{{\N}\kc}{\Jmax}.
\end{align}
Hence, the solution to Eq.~\ref{summary} depends on the four dimensionless parameters, $\kappa$, $\delta$, $\chib$, and $\chic$.

Notice that the parameter $\delta$ in Eq.~\ref{dp} compares the volume concentrations of receptors to substrates. The Briggs-Haldane analysis of Michaelis-Menten enzyme kinetics assumes that this parameter is small \cite{keener09}. In particular, if
\begin{align}\label{mmregime}
\delta\ll\kappa,
\end{align}
then we assume Eq.~\ref{mm} is in quasi-steady state,
\begin{align*}
0
%\approx{{{\kappa}}}\big(1-\c(\t)\big)\u(1, \t)-(\chi_{\text{b}}+\chi_{\text{c}})\c(\t),
\approx \u(1, \t)-\Big(\frac{\chi_{\text{b}}+\chi_{\text{c}}}{{{{\kappa}}}}+\u(1, \t)\Big)\c(\t),
\end{align*}
and thus,
\begin{align*}
\c(\t)
\approx\frac{\u(1,\t)}{(\chi_{\text{b}}+\chi_{\text{c}})/\kappa+\u(1,\t)}.
\end{align*}
In this parameter regime, the problem in Eq.~\ref{summary} becomes 
\begin{align}\label{summarymm}
\begin{split}
{\pt}u
&=D\Delta%\Big(\frac{2}{r}{\pr}+\frac{\partial^{2}}{\partial r^{2}}\Big)
u,\quad r>R,\,t>0,\\
\lim_{r\to\infty}u(r,t)
&=u_{0}>0,\\
D{\pr}u(R,t)
%&=\frac{\overline{V}_{\text{max}}\u(1,\t)}{\overline{K}_{\text{M}}+\u(1,\t)},
&=\frac{V u(R,t)}{K+u(R,t)},
\end{split}
\end{align}
where the maximum velocity and half-saturation constant in the boundary condition are
\begin{align}\label{k0}
V
:=e_{0}\kc=\frac{N}{4\pi R^{2}}\kc,\quad
K
:=\frac{\kb+\kc}{\ka}
=\frac{{\N}(\kb+\kc)}{4\pi D R \kappa}.
\end{align}
That is, the ODE boundary condition in Eq.~\ref{summary} is replaced by a Michaelis-Menten type boundary condition in Eq.~\ref{summarymm}.

%%%%%%%%%%%%%%%%%%%%%%%%%%%%%%%%%%%%%%%%%%%%%%%%%%%%%%%%%%%%%%%%%%%%%%%%%%%%%%%%%%%%%%%%%%%%%%%%%%%%%%%%%%%%
\subsubsection*{Steady-state uptake and receptor occupation}\label{sssection}

At steady-state, solving the full PDE-ODE system in Eq.~\ref{summary} is equivalent to solving the Michaelis-Menten system in Eq.~\ref{summarymm}. In either case, it is straightforward to obtain 
\begin{align}
u_{\ss}(r)
&:=\lim_{t\to\infty}u(r,t)
=u_{0}\Big(1-a\frac{R}{r}\Big),\nonumber\\
c_{\ss}
&:=\lim_{t\to\infty}c(t)
=\Big(\frac{1-a}{(\chib+\chic)/\kappa+1-a}\Big)e_{0},\label{css}
%=\frac{e_{0}{{{\kappa}}}(1-a)}{\chi_{\text{b}}+\chi_{\text{c}}+{{{\kappa}}}(1-a)}
%=\frac{ae_{0}}{\chic},
\end{align}
where $a$ is the dimensionless parameter,
\begin{align}\label{a}
a
=\frac{1}{2}\left(\chic+\frac{\chi_{\text{b}}+\chi_{\text{c}}}{{{{\kappa}}}}+1-\sqrt{\Big(\chic+\frac{\chi_{\text{b}}+\chi_{\text{c}}}{{{{\kappa}}}}+1\Big)^{2}-4\chic}\right).
%\in\Big(0,\frac{\kappa}{\kappa+1}\Big).
\end{align}
The fraction of receptors which are bound at steady-state is $c_{\ss}/e_{0}\in(0,1)$. The steady-state total flux into the cell is
\begin{align}\label{J}
\J
:=D\int_{r=R}{\dr}u_{\ss}\,\dd S
%=\chic\c_{\ss}\Jmax
%=4\pi DaRu_{0}
=aJ_{\text{max}}
%<\frac{{{{\kappa}}}}{{{{\kappa}}}+1}J_{\text{max}}=
<J_{\text{bp}}.
\end{align}

The inequality in Eq.~\ref{J} is the desired result that the flux into the cell when the receptors have finite kinetics is strictly less than the flux into the cell when the receptors process molecules at infinite kinetic rates. To verify Eq.~\ref{J}, note first that the case $\chic=0$ is trivial since $a=0$ if and only if $\chic=0$. Next, suppose $\chic>0$. Note that Eq.~\ref{bpg} means $J_{\text{bp}}=a_{\text{bp}}J_{\text{max}}$ where $a_{\text{bp}}:={{{\kappa}}}/({{{\kappa}}}+ 1)$. Hence, $a_{\text{bp}}$ satisfies
\begin{align}\label{e1}
a_{\text{bp}}
={{{\kappa}}}(1-a_{\text{bp}}).
\end{align}
On the other hand, the boundary condition at $r=R$ implies that $a$ satisfies
\begin{align}\label{e2}
a=\frac{{{{\kappa}}}(1-a)}{1+\chi_{\text{b}}/\chi_{\text{c}}+({{{\kappa}}}/\chi_{\text{c}})(1-a)}.
\end{align}
It is clear that the solution to Eq.~\ref{e1} is larger than the solution to Eq.~\ref{e2} since ${{{\kappa}}}$ and $\chic$ are strictly positive. Therefore,
\begin{align*}
a\in\Big(0,\frac{\kappa}{\kappa+1}\Big)=(0,a_{\text{bp}}),
\end{align*}
which verifies Eq.~\ref{J}.

In addition, fixing $\chib$ and ${{{\kappa}}}$ and taking $\chic\to\infty$ in Eq.~\ref{e2} and comparing to Eq.~\ref{e1} shows that
\begin{align}\label{ce}
\J
=aJ_{\text{max}}
\to a_{\text{bp}}J_{\text{max}}
=\Jbp\quad\text{as }\chic\to\infty.
\end{align}
That is, $\J$ reduces to $\Jbp$ if the receptor turnover rate $\kc$ is much faster than $\Jmax/{\N}$.

%%%%%%%%%%%%%%%%%%%%%%%%%%%%%%%%%%%%%%%%%%%%%%%%%%%%%%%%%%%%%%%%%%%%%%%%%%%%%%%%%%%%%%%%%%%%%%%%%%%%%%%%%%%%%%%%%%%%%%%%%%%%%%%%%%%%%%%%%%%%%%%%%%%%%%%%%%%%%%
\subsubsection*{Other kinetic schemes and receptor internalization}\label{gk}

The analysis above extends to more general kinetic schemes than the standard substrate-enzyme reaction in Eq.~\ref{reaction}. Indeed, alternative kinetic schemes merely yield different systems of ODEs at the cell surface.

To illustrate, suppose that receptors transport substrate molecules by endocytosis (i.e.\ receptor internalization), which is often seen in eukaryotic cells \cite{aquino2010, ferguson2001, mukherjee1997}. In this case, we replace Eq.~\ref{reaction} by
\begin{align*}
S+E\Markov{{\kb}}{{\ka}}C\xrightarrow{{\kc}} P,
\quad E\xrightarrow{\kc^{0}}\varnothing,
\quad \varnothing\xrightarrow{\kr} E,
\end{align*}
where $\kc$ and $\kc^{0}$ are the respective internalization rates for bound receptors $C$ and free receptors $E$, and $\kr$ is the rate that free receptors are delivered to the membrane. In this case, the boundary condition at $r=R$ in Eq.~\ref{bc11} for the substrate flux is unchanged and the ODEs in Eq.~\ref{ode} become 
\begin{align*}
{\dt}e(t)
&=-{\ka}e(t)u(R, t)+{\kb}c(t)-\kc^{0}e(t)+\kr, \\
{\dt}c(t)
&={\ka}e(t)u(R, t)-({\kb}+{\kc})c(t).
\end{align*}

%%%%%%%%%%%%%%%%%%%%%%%%%%%%%%%%%%%%%%%%%%%%%%%%%%%%%%%%%%%%%%%%%%%%%%%%%%%%%%%%%%%%%%%%%%%%%%%%%%%%%%%%%%%%%%%%%%%%%%%%%%%%%%%%%%%%%%%%%%%%%%%%%%%%%%%%%%%%%%%%%%%%%%%%%%%%%%%%%%%%%%%%%%%%%%%%%%%%%%%%%%%%%%%%%%%%%%%%%%%%%%%%%%%%%%%%%%%%%%%%%%%%%%%%%%%%%%%%%%%%%%%%%%%%%
\subsection*{Numerical methods and simulations}\label{numerics}

To verify the predictions of the modeling framework developed above, we perform numerical simulations of a stochastic, many particle system. To reduce computational cost, the stochastic simulations are performed in a cylindrical spatial domain. We begin by extending the analysis above to this spatial domain.

%%%%%%%%%%%%%%%%%%%%%%%%%%%%%%%%%%%%%%%%%%%%%%%%%%%%%%%%%%%%%%%%%%%%%%%%%%%%%%%%%%%%%%%%%%%%%%%%%%%%%%%%%%%%
\subsubsection*{Cylindrical domain}\label{geometry}

Let the spatial domain $\Omega$ be a cylinder of radius $2{R_{0}}>0$ and height $L>0$, 
\begin{align}\label{cyl}
\Omega
:=\big\{(x,y,z)\in\R^{3}:x^{2}+y^{2}<4{R_{0}}^{2},\,z\in(0,L)\big\}.
\end{align}
Substrate molecules diffuse in $\Omega$ with reflecting boundaries at the top ($z=L$) and the sides ($r:=\sqrt{x^{2}+y^{2}}=2{R_{0}}$). Hence, the substrate concentration $u=u(x,y,z,t)$ satisfies
\begin{align*}
{\pt}u
&=D\Delta u,\quad (x,y,z)\in\Omega,\,t>0,\\
{\pr}u
&=0,\quad r=2{R_{0}};\qquad
{\pz}u
=0,\quad z=L.
\end{align*}
Analogous to the surface of the sphere in the model above, we assume that the bottom of the cylinder ($z=0$) is reflecting, except for ${\N}\gg1$ small circular receptors. 

If the receptors process substrates continuously, then the substrate concentration satisfies mixed boundary conditions at the bottom of the cylinder (analogous to Eq.~\ref{bc}),
\begin{align}\label{bc2}
\begin{split}
D{\pz}u&=\frac{D}{{R_{0}}}\kappa_{\text{rec}}u,\quad \,z=0,\text{ and }(x,y)\text{ in a receptor},\\
{\pz}u&=0,\qquad z=0,\text{ and }(x,y)\text{ not in a receptor},
\end{split}
\end{align}
where $\kappa_{\text{rec}}$ is as in Eq.~\ref{kapparec}. If the $N$ receptors are roughly evenly distributed and have common radius $\eps {R_{0}}\ll {R_{0}}$, then Eq.~\ref{bc2} can be replaced by 
\begin{align}\label{zkappa}
D{\pz}u
&=\frac{D}{{R_{0}}}\kappa u,\quad z=0,
\end{align}
where $\kappa$ is in Eq.~\ref{trapping}. Hence, the problem reduces to a one-dimensional PDE for $u=u(z,t)$,
\begin{align}
{\pt}u
&=D\Delta u,\quad z\in(0,L),\,t>0,\label{zlap}\\
{\pz}u
&=0,\quad z=L,\label{urefl}
\end{align}
with the boundary condition in Eq.~\ref{zkappa} at $z=0$.

As above, we can incorporate finite receptor kinetics by replacing the boundary condition at $z=0$ by a boundary condition that couples to an ODE. Specifically,
\begin{align}
D{\pz}u(0,t)
&={\ka}\big(e_{0}-c(t)\big) u(0,t)-{\kb}c( t),\label{ucyl}\\
{\dt}c(t)
&={\ka}\big(e_{0}-c(t)\big)u(0, t)-({\kb}+{\kc})c(t),\label{ccyl}
\end{align}
where $c(t)$, $e_{0}$, $\ka$, $\kb$, and $\kc$ are as above (with $R$ replaced by $R_{0}$ in Eqs.~\ref{e0} and \ref{ka}). Furthermore, in the parameter regime in Eq.~\ref{mmregime}, we can eliminate Eq.~\ref{ccyl} and replace Eq.~\ref{ucyl} by a Michaelis-Menten type boundary condition with $V$ and $K$ given in Eq.~\ref{k0},
\begin{align}
D{\pz}u(0,t)
&=\frac{V u(0,t)}{K+u(0,t)}.\label{mmcyl}
\end{align}

%%%%%%%%%%%%%%%%%%%%%%%%%%%%%%%%%%%%%%%%%%%%%%%%%%%%%%%%%%%%%%%%%%%%%%%%%%%%%%%%%%%%%%%%%%%%%%%%%%%%%%%%%%%%%%%%%%%%%%%%%%%%%%%%%%%%%%%%%%%%
\subsubsection*{Stochastic simulation method}

We now describe the stochastic simulation method. We use the standard Euler-Maruyama method \cite{kloeden1992} for simulating the paths of many diffusing substrate molecules in $\Omega$ with reflecting boundary conditions on the boundaries away from receptors. If a molecule hits a ``free'' receptor, then that molecule immediately binds to the receptor (corresponding to $\kappa_{\text{rec}}=\infty$ in Eq.~\ref{kapparec}). If a receptor has a molecule bound to it, then that receptor is considered ``occupied'' and any other molecule that hits it simply reflects. We take $\kb=0$, and thus a bound molecule is removed from the system after an exponentially distributed time with rate $\kc>0$. When a bound molecule is removed from the system, the corresponding receptor changes from ``occupied'' back to ``free'' and can thus bind additional molecules. 

All stochastic simulations simulations were written in a combination of C and MATLAB~\cite{MATLAB}. The simulations were completed in the cylinder in Eq.~\ref{cyl} with height $L=1\,\mu\text{m}$ and radius $2{R_{0}}=0.1\,\mu\text{m}$, with $N=500$ receptors of common radius $0.001\,\mu\text{m}$ placed uniformly at random (non-overlapping) along the disk centered at $z = 0$. We take $D=10^{3}\,\mu\text{m}^{2}\text{s}^{-1}$ and $\kc\in\{10,10^{2},10^{3},10^{4}\}\,\text{s}^{-1}$. Each trial began with all receptors ``free'' and $10^{4}$ particles placed in the domain according to a normal distribution with mean $(x_{0},y_{0},z_{0})=(0,0,0.9)\,\mu\text{m}$ and standard deviation $0.01\,\mu\text{m}$ in each direction. For each value of $\kc$, 10 trials were completed with a discrete time step of $10^{-9}\,\text{s}$. Additional trials and smaller time steps were tested on a subset of simulations and did not yield significant quantitative changes.

%%%%%%%%%%%%%%%%%%%%%%%%%%%%%%%%%%%%%%%%%%%%%%%%%%%%%%%%%%%%%%%%%%%%%%%%%%%%%%%%%%%%%%%%%%%%%%%%%%%%%%%%%%%%%%%%%%%%%%%%%%%%%%%%%%%%%%%%%%%%%%%%%%%%%%%%%%%%%%%%%%%%%%%%%%%%%%%%%%%%%%%%%%%%%%%%%%%%%%%%%%%%%%%%%%%%%%%%%%%%
\subsubsection*{PDE numerical solution method}
We numerically solve the PDE-ODE system (Eqs.~\ref{zlap}, \ref{urefl}, and \ref{ucyl}-\ref{ccyl}) and the PDE with a Michaelis-Menten boundary condition (Eqs.~\ref{zlap}, \ref{urefl}, and \ref{mmcyl}) with the method of lines \cite{leveque2007}. Essentially, this method approximates the PDE with a large system of ODEs by replacing spatial derivatives with finite differences. The method is fairly standard, but the nonstandard boundary conditions must be handled carefully.

We now give the details of the method. We approximate $u(z,t)$ at ${{n}}\gg1$ off-center grid points,
\begin{align}
z_{j}:=(j+\tfrac{1}{2})\Delta z,\quad\text{for }j=0,1,\dots,{{n}}-1,
\end{align}
where $\Delta z=L/{{n}}\ll L$, and denote the approximation by $u_{j}(t)\approx u(z_{j},t)$. Replacing the Laplacian in Eq.~\ref{zlap} by a finite difference, $u_{j}(t)$ satisfies the ODE,
\begin{align}\label{so} 
\dt u_{j} =D\frac{(u_{j-1}-u_{j} +u_{j+1})}{(\Delta z)^2},\,\text{for }j=0,1,\dots,{{n}}-1.
\end{align}
Notice that the equations for $\dt u_0$ and $\dt u_{{{n}}-1}$ in Eq.~\ref{so} involve $u_{-1}$ and $u_{{{n}}}$, which are not yet defined. To ameliorate this issue, we make use of so-called ghost points, $z_{-1} := -\frac{1}{2}\Delta z$ and $z_{{{n}}} := ({{n}}+\frac{1}{2})\Delta z$, and solve for $u_{-1}$ and $u_{{{n}}}$ using the boundary conditions. Specifically, we approximate the boundary condition at $z=L$ in Eq.~\ref{urefl} with a finite difference,
\begin{align}\label{ghost0}
\frac{u_{{{n}}}-u_{{{n}}-1}}{\Delta z}
=0.
\end{align}
Hence, Eq.~\ref{ghost0} implies $u_{{{n}}}=u_{{{n}}-1}$, which we then use to solve Eq.~\ref{so} when $j={{n}}-1$.

In the case of the PDE-ODE system, we approximate the boundary condition at $z=0$ in Eq.~\ref{ucyl} by 
\begin{align}  \label{ghost1}
	D\frac{(u_{0}-u_{-1})}{\Delta z} &= \ka  (e_{0}-c(t)) \Big(\frac{u_0 + u_{-1}}{2}\Big) - \kb c(t),
\end{align}
where we have replaced $u(0,t)$ by $(u_{0}+u_{-1})/2$. We then solve Eq.~\ref{ghost1} for $u_{-1}$ and use this in Eq.~\ref{so} when $j=0$. In addition to the ${{n}}$ ODEs in Eq.~\ref{so}, we also have the ODE for $c(t)$ which is obtained from Eq.~\ref{ccyl} upon replacing $u(0,t)$ by $(u_{0}+u_{-1})/2$. 

In the case of the Michaelis-Menten boundary condition, we approximate Eq.~\ref{mmcyl} by 
\begin{align}  \label{ghost2}
	D\frac{(u_{0}-u_{-1})}{\Delta z} &= \frac{V(u_{0} + u_{-1})/2}{K+(u_{0} + u_{-1})/2}.
\end{align}
We then solve Eq.~\ref{ghost2} for $u_{-1}>0$ and use this in Eq.~\ref{so} when $j=0$.

Summarizing, the PDE-ODE system is approximated by ${{n}}+1$ ODEs, and the PDE with a Michaelis-Menten boundary condition is approximated by ${{n}}$ ODEs. In either case, the ODEs are solved using the \texttt{ode15s} function in MATLAB \cite{MATLAB} with $n=10^{3}$ spatial grid points.

%\subsubsection*{Code avaliability}
%Code for reproducing the stochastic and PDE numerical simulations are avaliable on the GitHub database (https://github.com/ gregoryhandy).

%%%%%%%%%%%%%%%%%%%%%%%%%%%%%%%%%%%%%%%%%%%%%%%%%%%%%%%%%%%%%%%%%%%%%%%%%%%%%%%%%%%%%%%%%%%%%%%%%%%%%%%%%%%%%%%%%%%%%%%%%%%%%%%%%%%%%%%%%%%%%%%%%%%%%%%%%%%%%%%%%%%%%%%%%%%%%%%%%%%%%%%%%%%%%%%%%%%%%%%%%%%%%%%%%%%%%%%%%%%%%%%%%%%%%%%%%%%%%%%%%%%%%%%%%%%%%%%%%%%%%%%%%%%%%%%%%%%%%%%%%%%%%%%%%%%%%%%%%%%%%%%%%%%%%%%%%%%%%%%%%%%%%%%%%%%%%%%%%%%%%%%%%%%%%%%%%%%%%%%%%%%%%%%%%%%%%
\section*{Results and Discussion}

%%%%%%%%%%%%%%%%%%%%%%%%%%%%%%%%%%%%%%%%%%%%%%%%%%%%%%%%%%%%%%%%%%%%%%%%%%%%%%%%%%%%%%%%%%%%%%%%%%%%%%%%%%%%%%%%%%%%%%%%%%%%%%%%%%%%%%%%%%%%%%%%%%%%%%%%%%%%%%%%%%%%%%%%%%%%%%%%%%%%%%%%%%%%%%%%%%%%%%%%%%%%%%%%%%%%%%
\subsection*{Analysis confirmed by stochastic simulations}

\begin{figure}[t]%[htp][hbt!]%
	\centering
\includegraphics[width=.6\linewidth]{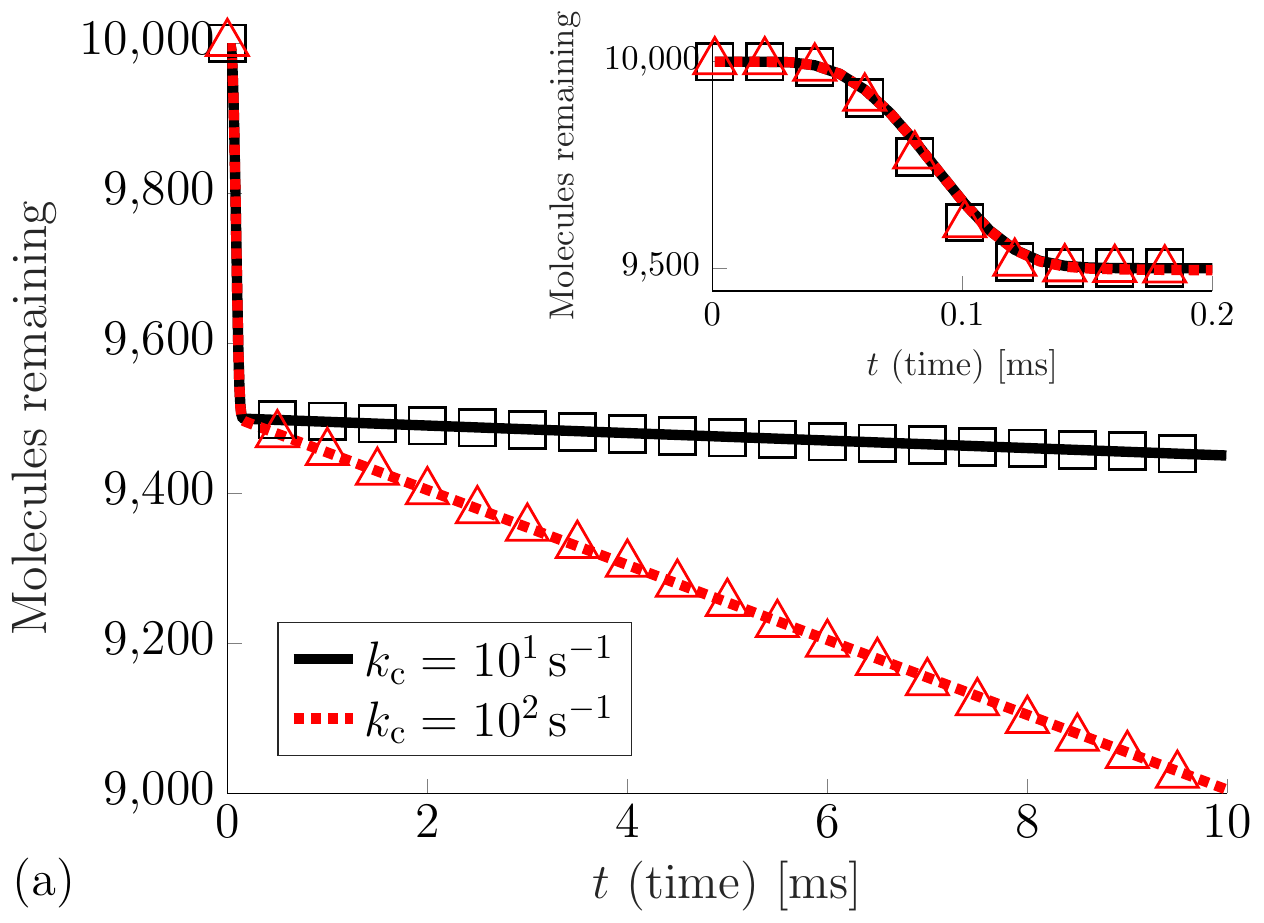}
\includegraphics[width=.6\linewidth]{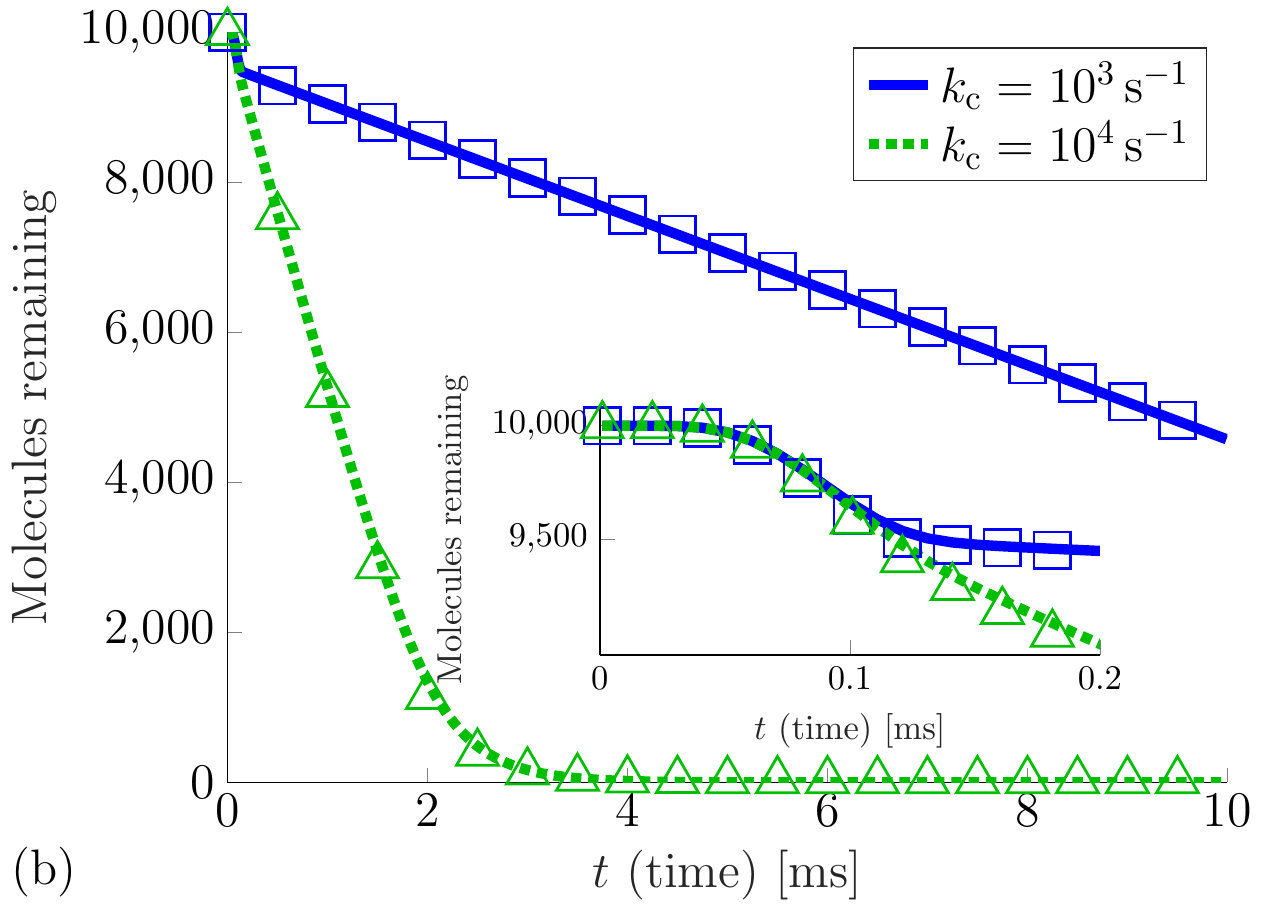}
	\caption{
Comparison of stochastic simulations (squares and triangles) to the deterministic PDE-ODE system (solid and dotted curves) for the cylindrical domain for different values of the receptor turnover rate $\kc$. The insets zoom in at early times. See the text for details.
	}
	\label{figsim}
\end{figure}

In Fig.~\ref{figsim}, we plot the results of stochastic simulations (squares and triangles) and the solution to the deterministic PDE-ODE system (solid and dotted curves) for the cylindrical spatial domain in Eq.~\ref{cyl} for various choices of the receptor turnover rate $\kc$. Specifically, we plot the number of diffusing molecules remaining in the domain (molecules which are unbound and have not been absorbed) as a function of time. This plot shows that the PDE-ODE system accurately describes the dynamics of the full stochastic system involving many interacting particles.

Notice in Fig.~\ref{figsim} that at early times ($t<0.2\,\text{ms}$), the number of remaining molecules rapidly drops from $10,000$ to $9,500$, which corresponds to $500$ molecules quickly binding the $N=500$ receptors. This initial rapid decrease is seen in both the PDE-ODE solution and the stochastic simulations, which is evident from the insets in Fig.~\ref{figsim} which zoom in at early times. Then as time increases, the number of remaining molecules decreases linearly with a slope of $N\kc$, which is readily seen in Fig.~\ref{figsim} for both the PDE-ODE solution and the stochastic simulations.

The PDE solution with a Michaelis-Menten boundary condition also produces the desired linear decay with slope $N\kc$, but it does not exhibit the initial rapid decay of $N$ molecules binding to the $N$ receptors (plot not shown). This is not surprising, since the Michaelis-Menten boundary condition was derived assuming that there are many more diffusing molecules than receptors per some characteristic volume. Indeed, in such a parameter regime, the size of the initial drop in the number of molecules (namely the number of receptors, $N$) would be small compared to the number of molecules and would thus be negligible.

To reduce computational expense, the simulations were performed in a bounded, cylindrical spatial domain rather the unbounded domain exterior to a sphere in Eq.~\ref{pde11}. However, we expect that the agreement between stochastic simulations and the PDE-ODE framework seen here extends to more general geometries, including the unbounded spherical geometry of Eq.~\ref{pde11}. In fact, the cylindrical domain in Eq.~\ref{cyl} can model a cylindrical region of height $L=1\,\mu\text{m}$ and radius $2R_{0}=0.1\,\mu\text{m}$ directly above a cell, where the base of the cylinder represents a flat patch of cell membrane with many receptors. In particular, for cells of radius $R\ge1\,\mu\text{m}$, the membrane curvature is negligible at the base of a cylinder of radius $2R_{0}=0.1\,\mu\text{m}$.

%%%%%%%%%%%%%%%%%%%%%%%%%%%%%%%%%%%%%%%%%%%%%%%%%%%%%%%%%%%%%%%%%%%%%%%%%%%%%%%%%%%%%%%%%%%%%%%%%%%%%%%%%%%%%%%%%%%%%%%%%%%%%%%%%%%%%%%%%%%%%%%%%%%%%%%%%%%%%%%%%%%%%%%%%%%%%%%%%%%%%%%%%%%%%%%%
\subsection*{Parameter ranges}\label{params}

%%%%%%%%%%%%%%%%%%%%%%%%
\begin{table}[t]
\begin{center}
\footnotesize
%\small
\begin{center}
\begin{tabular}{ | lcr | } 
 \hline
  \rule{0pt}{2.25ex} 
 Parameter & Default value & Range of interest \\
 \hline
 \hline
\rule{0pt}{2.25ex}    
$D=$  molecule diffusivity & $10^{3}\,\mu\text{m}^{2}\text{s}^{-1}$ & $[10^{2},10^{3}]\,\mu\text{m}^{2}\text{s}^{-1}$\\ 
\hline
 \rule{0pt}{2.25ex}    
$R=$  cell radius & $1\,\mu\text{m}$ & $[0.35,15]\,\mu\text{m}$\\
\hline
\rule{0pt}{2.25ex}    
$\eps=$  receptor to cell radius ratio & $10^{-3}$ & \\ 
\hline
\rule{0pt}{2.25ex}    
${\N}=$  number of receptors & $10^{3}$ & $[10^{2},10^{5}]$ \\ 
\hline
\rule{0pt}{2.25ex}    
$\kappa_{\text{rec}}=$  receptor trapping rate & $\infty$ & \\ 
\hline
\rule{0pt}{2.25ex}    
$u_{0}=$  extracellular concentration & $1\,\mu\text{M}$ & $[1\,\text{nM},1\,\text{mM}]$\\ 
\hline
\rule{0pt}{2.25ex}    
$\kc=$  receptor turnover rate & $10^{2}\,\text{s}^{-1}$ & $[10^{1},10^{5}]\,\text{s}^{-1}$ \\ 
\hline
\rule{0pt}{2.25ex}    
$\kb=$  receptor unbinding rate & $0$ & $[0,10^{4}]\,\text{s}^{-1}$ \\ 
 \hline
\end{tabular}
\end{center}
\normalsize
\end{center}
\caption{
Summary of parameter values and ranges. Unless otherwise noted, the ``default'' values are used in the figures and calculations. See the text for details.
}
\label{tableparams}
\end{table}
%%%%%%%%%%%%%%%%%%%%%%%%

Before discussing some biophysical implications of our analysis, we briefly discuss parameter values. We do not seek precise values for any one specific application, but rather we choose ranges and orders of magnitude that are relevant across multiple systems. Unless otherwise noted, the following ``default'' parameter values are used in the figures and calculations below. The parameters are summarized in Table~\ref{tableparams}.

Cell radii range between roughly $0.35\,\mu\text{m}$ for a small bacteria to $15\,\mu\text{m}$ for a large mammalian cell \cite{bionumbers}. We set the default radius to be $R=1\,\mu\text{m}$, which is consistent with a bacterial cell or a small eukaryotic cell. We set the default diffusion coefficient to be $D=10^{3}\,\mu\text{m}^{2}\text{s}^{-1}$, which is the order of magnitude relevant for glucose uptake by an \emph{E.\ coli} cell or a yeast cell \cite{meijer1996, maier2002, natarajan1999, lavrentovich2013} and chemotaxis by bacterial cells and slime mold \cite{berg1977}. Following \cite{berg1977, wagner2006}, we take the radius of each receptor to be $\eps R=1\,\text{nm}$ and $\kappa_{\text{rec}}=\infty$, which means $\eps=10^{-3}$ and $\kappa=\eps {\N}/\pi$. The number of receptors ${\N}$ on a cell can vary greatly \cite{perelson1997, ismael2016, lawley2020prl}, and we take ${\N}=10^{3}$ as the default value. Extracellular concentrations of interest also vary considerably. For example, a characteristic value in \cite{berg1977} is $u_{0}=1\,\mu\text{M}$, the nutrient uptake study~\cite{wagner2006} considers $u_{0}=100\,\mu\text{M}$, and other nutrient uptake studies involve $u_{0}$ on the order of $10^{3}\,\mu\text{M}$ or greater \cite{meijer1996, maier2002, lavrentovich2013}. We follow \cite{berg1977} and take $u_{0}=1\,\mu\text{M}$ as the default value.

Finally, the kinetic rate parameters $\kb$ and $\kc$ can also vary considerably. The typical turnover rate for sugar transporters is $\kc=10^{2}\,\text{s}^{-1}$, with a range of $\kc\in[3\times10^{1}\,\text{s}^{-1},3\times10^{2}\,\text{s}^{-1}]$, though chloride-bicarbonate transporters can reach speeds on the order of $\kc=10^{5}\,\text{s}^{-1}$ \cite{bionumbers}. We take $\kc=10^{2}\,\text{s}^{-1}$ as the default value. Breakup rates $\kb$ have been estimated on the order of $10^{-4}\,\text{s}^{-1}$ \cite{aquino2010}, $10^{-3}\,\text{s}^{-1}$ \cite{mugler2016, aquino2010}, $1\,\text{s}^{-1}$ \cite{mugler2016}, and $10^{4}\,\text{s}^{-1}$ \cite{berg1977}. Since most values satisfy $\kb\ll\kc$, we take $\kb=0$ as the default value for simplicity.

%%%%%%%%%%%%%%%%%%%%%%%%%%%%%%%%%%%%%%%%%%%%%%%%%%%%%%%%%%%%%%%%%%%%%%%%%%%%%%%%%%%%%%%%%%%%%%%%%%%%%%%%%%%%%%%%%%%%%%%%%%%%%%%%%%%%%%%%%%%%%%%%%%%%%%%%%%%%%%%%%%%%%%%%%%%%%%%%%%%%%%%%%%%%%%%%%%%%%%%%%%%%%%%%%%%%%%
\subsection*{Receptor kinetics can dominate uptake}

%%%%%%%%%%%%%%%%%%%%%
\begin{figure}[t]%[htp][hbt!]%
\centering
\includegraphics[width=.6\linewidth]{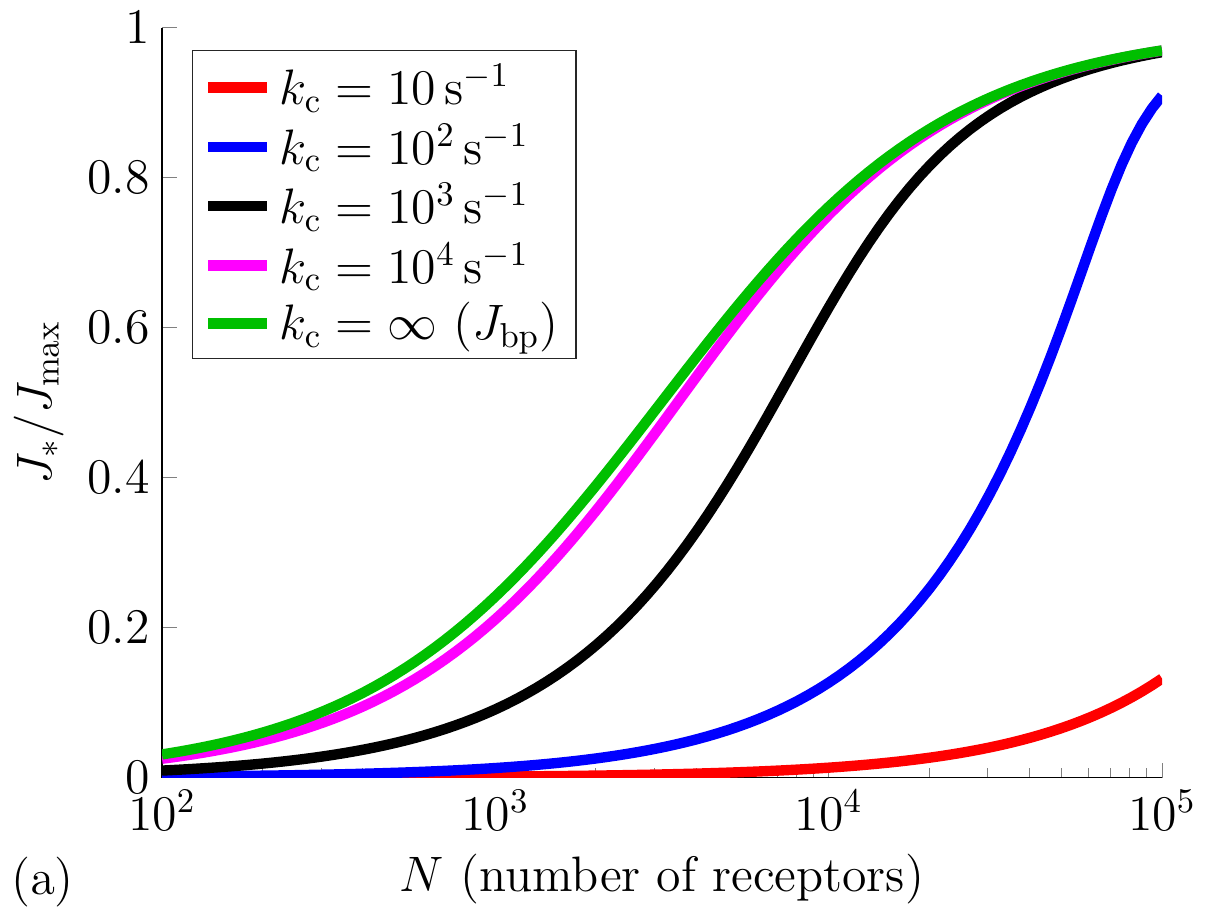}
\includegraphics[width=.6\linewidth]{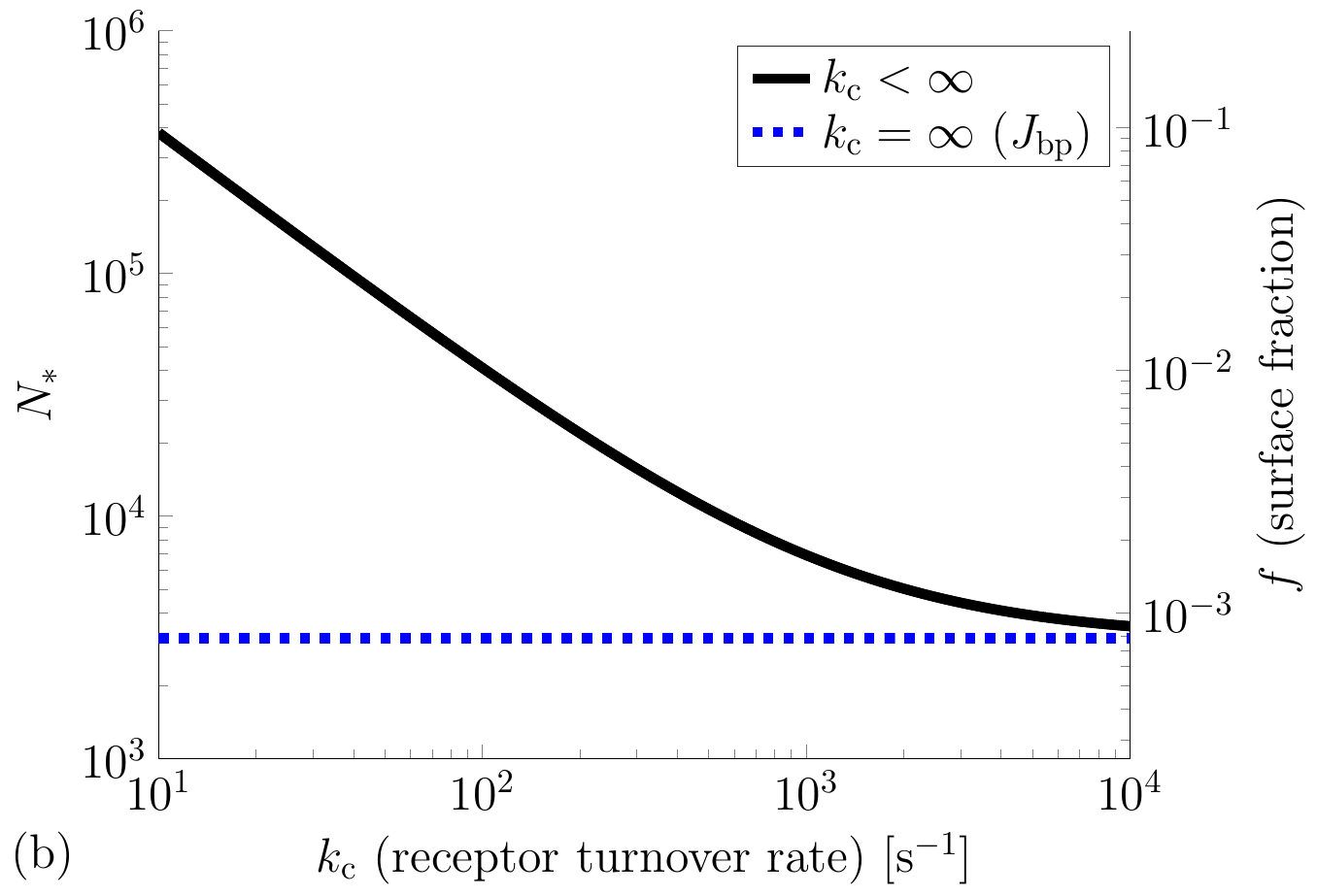}
\caption{(a) Cellular uptake as a function of the number of cell surface receptors for different turnover rates $\kc$. (b) Number of receptors needed so that $\J=\Jbp$ ($\Nhalf$ in Eq.~\ref{n12}) on the left axis as a function of $\kc$. The right axis gives the corresponding fraction of the cell surface covered by receptors ($f$ in Eq.~\ref{fraction}). See Table~\ref{tableparams} for parameter values.}
\label{figjm}
\end{figure}
%%%%%%%%%%%%%%%%%%%%%

In this subsection, we use our uptake formula ($\J$ in Eqs.~\ref{a}-\ref{J}) to show that finite receptor kinetics play a dominant role in cellular uptake in some typical biophysical scenarios. In Fig.~\ref{figjm}a, we plot the ratio $\J/\Jmax$ as a function of the number of receptors $N$ for different values of the receptor turnover rate $\kc$. Note that $\J$ reduces to the Berg-Purcell flux, $\Jbp$, if $\kc$ is infinite (see Eq.~\ref{ce}). This figure shows that $\J$ is much less than $\Jbp$ for typical values of the receptor turnover rate $\kc$. For example, if $\kc=10^{2}\,\text{s}^{-1}$ and the rest of the parameters are the default values in Table~\ref{tableparams}, then
\begin{align}\label{sd}
\frac{\J}{\Jbp}
\approx0.05.
\end{align}

The reason for the significant discrepancy in Eq.~\ref{sd} is that a large fraction of molecules that hit a receptor are blocked from binding because that receptor is occupied by another molecule (this is incorporated into $\J$ but ignored by $\Jbp$). Indeed, at these default parameter values, Eq.~\ref{css} implies that more than $95\%$ of the surface receptors are bound to a molecule at any given time. Hence, any molecule that manages to hit a receptor has less than a $5\%$ chance of binding upon first contact with that receptor.

It is also evident from Fig.~\ref{figjm}a that $\Jbp$ saturates at much smaller values of ${\N}$ compared to $\J$. For example, increasing the receptor number from ${\N}=10^{4}$ to ${\N}=10^{5}$ increases $\Jbp$ by a mere $27\%$, whereas this increase in the receptor number increases $\J$ by more than $600\%$ if $\kc=10^{2}\,\text{s}^{-1}$. This is because in the calculation of $\Jbp$, increasing the number of receptors merely increases the likelihood that a single molecule hits a receptor rather than escaping to spatial infinity. In contrast, if we include finite receptor kinetics, then increasing the number of receptors also increases the number of molecules that can be bound to the cell at any one time. 

To further investigate this point, in Fig.~\ref{figjm}b we plot on the left axis the number of receptors, $\Nhalf$, required for $\J$ to reach one half of $\Jmax$ as a function of $\kc$. That is, we plot
\begin{align}\label{n12}
\Nhalf
:=\frac{\pi}{\eps}+\frac{\Jmax}{2\kc}+\frac{\pi\kb}{\kc},
\end{align}
where the formula in Eq.~\ref{n12} was found by setting $\J=\Jmax/2$ and solving for $N$. The corresponding number of receptors required for the Berg-Purcell flux, $\Jbp$, to reach $\Jmax/2$ is
\begin{align}\label{nbp}
%\Nbp:=
\lim_{\kc\to\infty}\Nhalf
=\frac{\pi}{\eps}.
\end{align}
Using Eqs.~\ref{n12}-\ref{nbp}, we see that $\Jbp$ reaches one half of $\Jmax$ with $\Nhalf\approx3\times10^{3}$ receptors ($\kc=\infty$), whereas $\J$ requires $\Nhalf\approx4\times10^{4}$ receptors to reach one half of $\Jmax$ if $\kc=10^{2}\,\text{s}^{-1}$.

On the right axis of Fig.~\ref{figjm}b, we plot the corresponding fraction of the cell surface covered by receptors,
\begin{align}\label{fraction}
f
:=\frac{\pi(\eps R)^{2}\Nhalf}{4\pi R^{2}}
=\frac{\eps^{2}\Nhalf}{4}\in(0,1).
\end{align}
Interestingly, $f$ is very small as long as $\kc$ is not very slow. For example, $f\approx10^{-3}=0.1\%$ for $\Nhalf=3\times10^{3}$, and $f=10^{-2}=1\%$ for $\Nhalf=4\times10^{4}$. Therefore, the remarkable result of Berg and Purcell that a cell requires only a small receptor surface fraction $f$ in order to have uptake near the maximum $\Jmax$ still holds in the case of finite receptor kinetics.

As mentioned in the Introduction, many previous works have sought to modify and refine the Berg-Purcell formula to incorporate various details in the problem \cite{zwanzig1991, bernoff2018b, lindsay2017,Berezhkovskii2004,Berezhkovskii2006,Dagdug2016,Eun2017,Muratov2008,eun2020,lawley2019bp,kaye2019}. It is therefore worth pointing out that the discrepancy between $\J$ and $\Jbp$ is much greater than some previous modifications of $\Jbp$. For example, Zwanzig posited the formula \cite{zwanzig1990},
\begin{align*}
J_{\text{zw}}
:=\frac{\eps {\N}}{\eps {\N}+(1-\eps^{2}{\N}/4)\pi}\Jmax,
\end{align*}
in order to account for the effects of interference between receptors. However, the percentage difference between $\Jbp$ and $J_{\text{zw}}$ is less than the dimensionless receptor radius $\eps$ \cite{lawley2019bp}, 
\begin{align*}
\frac{J_{\text{zw}}-\Jbp}{J_{\text{zw}}}
=\frac{{\N}\pi\eps^{2}}{4{\N}\eps+4\pi}
\le\lim_{{\N}\to\infty}\frac{{\N}\pi\eps^{2}}{4{\N}\eps+4\pi}
\le\eps\frac{\pi}{4}<\eps.
\end{align*}
That is, $J_{\text{zw}}$ and $\Jbp$ typically differ by around one tenth of one percent, whereas $\J$ and $\Jbp$ can differ by at least an order of magnitude in typical biophysical scenarios. 

%%%%%%%%%%%%%%%%%%%%%%%%%%%%%%%%%%%%%%%%%%%%%%%%%%%%%%%%%%%%%%%%%%%%%%%%%%%%%%%%%%%%%%%%%%%%%%%%%%%%%%%%%%%%%%%%%%%%%%%%%%%%%%%%%%%%%%%%%%%%%%%%%%%%%%%%%%%%%%%%%%%%%%%%%%%%%%%%%%%%%%%%%%%%%%%%%%%%%%%%%%%%%%%%%%%%%%
\subsection*{Uptake is almost Michaelis-Menten}

%%%%%%%%%%%%%%%%%%%%%
\begin{figure}[t]%[htp][hbt!]%
\centering
\includegraphics[width=.6\linewidth]{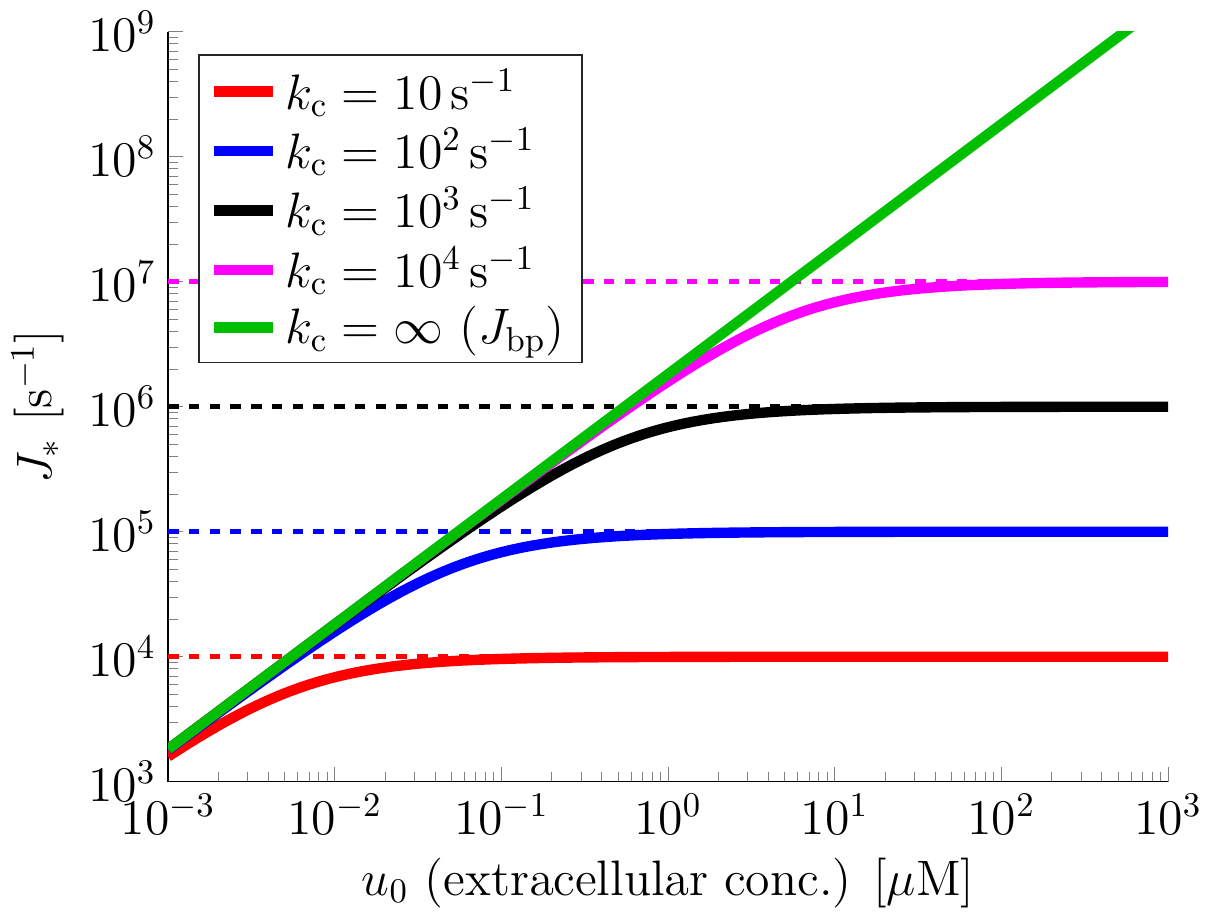}
\caption{Cellular uptake as a function of the extracellular concentration for different turnover rates $\kc$. The dashed horizontal lines are the maximum uptake rates for different turnover rates. See Table~\ref{tableparams} for parameter values.}
\label{figju1}
\end{figure}
%%%%%%%%%%%%%%%%%%%%%

The discrepancy between $\J$ and $\Jbp$ vanishes at low extracellular concentrations and is magnified at high extracellular concentrations. In Fig.~\ref{figju1}, we plot $\J$ and $\Jbp$ as functions of the extracellular concentration $u_{0}$. Here, we see that $\J$ and $\Jbp$ are indistinguishable at low concentrations (even for slow $\kc$ values), whereas $\J$ can be several orders of magnitude less than $\Jbp$ at high concentrations. The explanation for this is straightforward. At sufficiently low concentrations, molecules arrive at receptors at a much slower rate than receptors can process molecules, which is the assumption used to derive $\Jbp$ \cite{berg1977}. Since the per receptor arrival rate is $\Jbp/{\N}$ and the receptor kinetic rates are $\kb$ and $\kc$, a sufficiently low concentration is defined by the following dimensionless number being much less than one,
\begin{align}\label{sufflow}
\rho:=\frac{\Jbp/{\N}}{\kb+\kc}
=\frac{\frac{\kappa}{\kappa+1}4\pi DRu_{0}}{{\N}(\kb+\kc)}\ll1.
\end{align}
As the concentration increases and Eq.~\ref{sufflow} is violated, the arrival rate exceeds the receptor kinetic rates, and thus receptor kinetics modify cellular uptake. 

It is intuitively clear that cellular uptake cannot increase linearly as a function of $u_{0}$ indefinitely, but rather uptake must saturate at some maximum rate. In light of this observation, a Michaelis-Menten functional form for the uptake rate is often posited \cite{fiksen2013},
\begin{align}\label{jmm}
\Jmm
:=\frac{\Vmax u_{0}}{\KM+u_{0}},
\end{align}
for some maximum uptake rate $\Vmax$ and half-saturation constant $\KM$. We stress that the uptake equation in Eq.~\ref{jmm} is not to be confused with the boundary condition in Eq.~\ref{summarymm} which was derived to approximate the PDE-ODE system in Eq.~\ref{summary}.

The values of $\Vmax$ and $\KM$ in Eq.~\ref{jmm} are usually determined by matching Eq.~\ref{jmm} to experimental data. However, it is possible to relate $\Vmax$ and $\KM$ to microscopic biophysical parameters. First, it is clear that the maximum uptake rate must be $\Vmax={\N}\kc$. Next, if we force $\Jmm$ to coincide with $\Jbp$ at low concentrations (i.e.\ $u_{0}\ll\KM$), then we must have 
\begin{align}\label{kmj}
\KM
=\frac{\Vmax u_{0}}{\Jbp}
=\frac{{\N}\kc(1+\kappa)}{4\pi D R\kappa}.%,
\end{align}

Now, it is straightforward to use Eqs.~\ref{a}-\ref{J} to check that $\J$ has the desired property that it saturates at ${\N}\kc$ at high concentrations,
\begin{align*}
\lim_{u_{0}\to\infty}\J={\N}\kc,
\end{align*}
and $\J$ reduces to $\Jbp$ at low concentrations,
\begin{align*}
\lim_{u_{0}\to0}\J/\Jbp=1.
\end{align*}
Hence, $\J$ and $\Jmm$ agree at high and low concentrations. However, it is evident from the formula for $\J$ in Eqs.~\ref{a}-\ref{J} that $\J\neq\Jmm$ at intermediate concentrations.

%%%%%%%%%%%%%%%%%%%%%
\begin{figure}[t]%[htp][hbt!]%
\centering
\includegraphics[width=.6\linewidth]{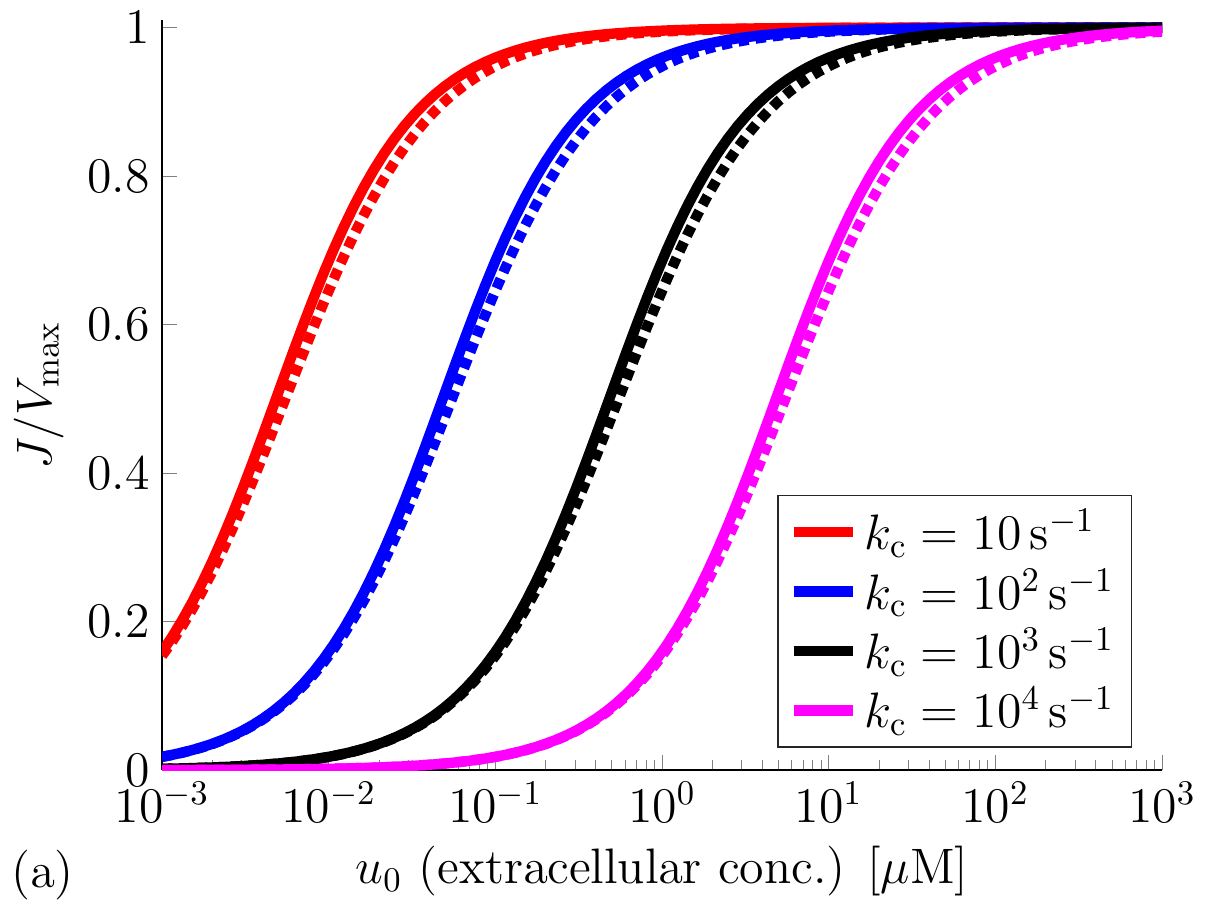}
\includegraphics[width=.6\linewidth]{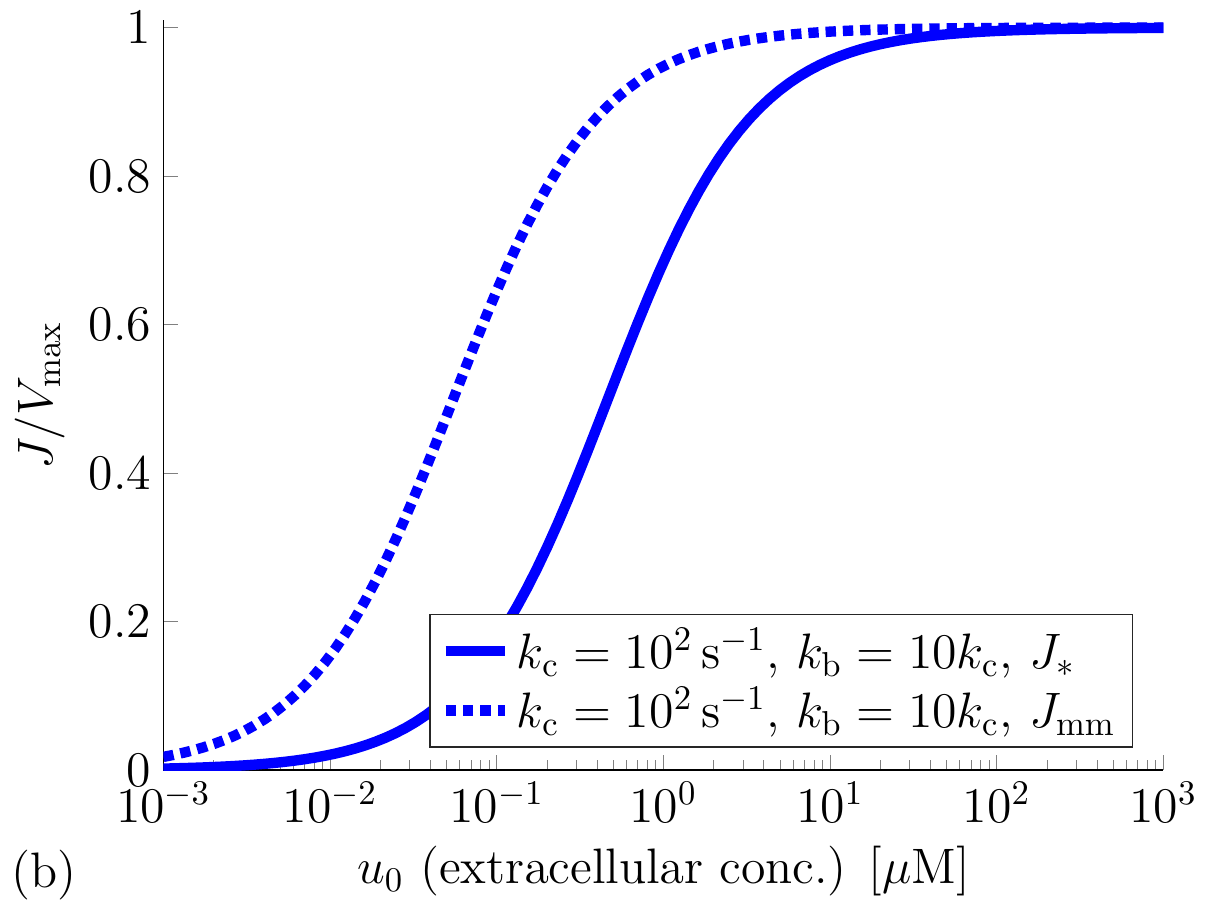}
\caption{Cellular uptake as a function of the extracellular concentration for different turnover rates $\kc$, with breakup rate $\kb=0$ in panel (a) and $\kb=10\kc$ in panel (b). The solid curves correspond to $\J$ in Eqs.~\ref{a}-\ref{J} and the dotted curves correspond to $\Jmm$ in Eq.~\ref{jmm}. See Table~\ref{tableparams} for other parameter values.}
\label{figju2}
\end{figure}
%%%%%%%%%%%%%%%%%%%%%

In Fig.~\ref{figju2}a, we plot $\J$ (solid curves) and $\Jmm$ (dotted curves) as functions of the concentration for $\kb=0$. While $\J$ does not have the exact Michaelis-Menten functional form as in Eq.~\ref{jmm}, this plot shows that the $\J$ curve does have a profile very similar to the profile generated by the Michaelis-Menten functional form (i.e.\ a sigmoidal curve). This is an important feature of the formula for $\J$, since this profile is observed in experiments measuring cellular uptake \cite{fiksen2013, meijer1996, maier2002, natarajan1999}.

Furthermore, it is straightforward to use the formula in Eqs.~\ref{a}-\ref{J} for $\J$ to find the half-saturation constant (i.e.\ the ``apparent $\KM$'' of $\J$). Indeed, solving the equation $\J=\Vmax/2$ for the concentration $u_{0}$ gives the half-saturation constant,
\begin{align}\label{u12}
%u_{\frac{1}{2}}
\KM^{*}
:=\frac{{\N}\kc(1+\kappa/2)}{4\pi DR\kappa}+\frac{{\N}\kb}{4\pi DR\kappa}.
\end{align}
Comparing Eq.~\ref{u12} with Eq.~\ref{kmj}, we see that these two half-saturation constants are similar if $\kb\ll\kc$. Indeed, the close agreement between $\J$ and $\Jmm$ in Fig.~\ref{figju2}a results from taking $\kb=0$. However, it follows from comparing Eq.~\ref{u12} and Eq.~\ref{kmj} that taking $\kb\not\ll\kc$ can make $\Jmm$ saturate at much lower concentrations than $\J$. This is illustrated in Fig.~\ref{figju2}b, where we plot $\J$ and $\Jmm$ with $\kb=10\kc$.

\section*{Conclusion}

We have developed a framework for modeling molecular species which diffuse in a three-dimensional bulk region and interact with receptors embedded on a two-dimensional surface. The receptors bind and process molecules at finite kinetic rates, which introduces significant statistical correlations between the individual diffusing molecules. We developed the framework in the context of the Berg-Purcell cellular uptake model \cite{berg1977}. We found that in some typical biophysical scenarios of interest, finite receptor kinetics can reduce cellular uptake by at least an order of magnitude compared to the Berg-Purcell estimate. The predictions of our analysis were confirmed by numerical simulations of a many particle stochastic system.

Mathematically, the framework uses the theory of boundary homogenization to couple a PDE (the diffusion equation) to nonlinear ODEs on a boundary. In a certain parameter regime (or at steady-state), the boundary conditions can be reduced to a nonlinear, Michaelis-Menten flux condition. Several interesting prior works have used PDEs with ODE boundary conditions to model reaction-diffusion systems \cite{levine2005, gomez2007, gou2016, gou2017, gomez2019, david2020, pb6}. These prior works have generally studied Turing patterns and spatiotemporal oscillations. To estimate how receptor diffusion and cellular rotation influence cellular uptake, Ref.~\cite{lawley2019bp} used the diffusion equation with boundary conditions described by stochastic differential equations. 

Previous works have employed mathematical models to study how finite receptor kinetics affect diffusive uptake. Refs.~\cite{handy2018, handy2019} formulated and analyzed stochastic models of diffusive interactions with receptors that must wait a transitory ``recharge'' time between captures.  In Ref.~\cite{handy2018}, it was proven that such a recharge time can drastically reduce the number of captured molecules (the number of captures grows logarithmically in the number of total molecules versus linear growth in the absence of recharge). Ref.~\cite{handy2019} used a variety of stochastic models to analyze similar systems. There is a rather large literature on cellular nutrient uptake which takes the Michaelis-Menten uptake equation in Eq.~\ref{jmm} as its starting point  \cite{fiksen2013, meijer1996, maier2002, natarajan1999}. The maximum uptake rate and half-saturation constant (i.e.\ $\Vmax$ and $\KM$) are chosen to fit experimental uptake rates. Our uptake formula ($\J$ in Eqs.~\ref{a}-\ref{J}) does not have the Michaelis-Menten functional form in Eq.~\ref{jmm}, but it nevertheless exhibits the same sigmoidal uptake curve as a function of the extracellular concentration.

While we developed the framework in the context of cellular uptake in a spherical geometry, it can be applied to other systems with potentially different geometries and receptor kinetic schemes. For example, synaptic transmission involves neurotransmitter molecules diffusing across the synaptic cleft and binding to receptors on the adjacent neuron \cite{deutch2013}. In this case, the shape of the synaptic cleft is similar to a cylinder and the framework of the present paper could be used to investigate the effect of the finite kinetic rates of neural receptors. As another example amenable to the present framework, cylindrical domains with receptors on the ``sides'' have been used to model catheter-based drug delivery systems \cite{hossain2012}.

Finally, we conclude by discussing the important study by Wagner et al.~\cite{wagner2006} on nutrient uptake by bacterial cells. This work used fluorescent tracing to show that bacterial cell ``stalks,'' which are long and thin extensions of the cell envelope, can bind and import nutrients from the extracellular environment. These authors then used novel mathematical analysis to generalize the Berg-Purcell model to a domain exterior to a stalk (modeled by a prolate spheroid). Based on this analysis, the authors argued that the stalk morphology increases nutrient uptake compared to a sphere. For future work, it would be interesting to extend the analysis in the present paper to the geometry considered in \cite{wagner2006} to investigate the effect of finite receptor kinetics.

Indeed, the model in \cite{wagner2006} assumes that receptors can absorb nutrient molecules continuously. As in Berg-Purcell, this assumption is valid if receptors can import molecules at a much faster rate than molecules tend to arrive at receptors. We now use the parameter $\rho$ in Eq.~\ref{sufflow} to compare these two rates in \cite{wagner2006}, where $\rho\ll1$ means that receptor kinetics are much faster than arrivals. The model in \cite{wagner2006} considered $N=10^{4}$ perfectly absorbing receptors (meaning $\kappa_{\text{rec}}=\infty$ in our notation). The radius of each receptor was $1\,\text{nm}$ and stalk lengths varied from $1\,\mu\text{m}$ to $10\,\mu\text{m}$. The extracellular concentration used in the experiments was $u_{0}=100\,\mu\text{M}$. If we take these values and set $R=1\,\mu\text{m}$, $D=10^{3}\,\mu\text{m}^{2}s^{-1}$, and $\kb=0$ in Eq.~\ref{sufflow}, then the receptor turnover rate $\kc$ would need to satisfy
\begin{align*}
\kc\gg5.7\times10^{4}\,\text{s}^{-1}
\end{align*}
in order to have $\rho\ll1$. That is, a single receptor would need to import molecules at a rate much faster than $10^{4}\,\text{s}^{-1}$, whereas characteristic rates are on the order of $10^{2}\,\text{s}^{-1}$ \cite{bionumbers}. While this simple calculation ignores the stalk geometry, it nevertheless suggests that finite receptor kinetics may play an important role in the uptake rate for the system studied in Ref.~\cite{wagner2006}. More broadly, the framework developed in the present paper provides a method for investigating how receptor kinetics affect a variety of biophysical systems.

\section*{Acknowledgments}

The first author was supported by The Swartz Foundation. The second author was supported by the National Science Foundation (Grant Nos.\ DMS-1944574, DMS-1814832, and DMS-1148230).

% Create the reference section using BibTeX:
\bibliography{library.bib}
\bibliographystyle{unsrt}

\end{document}